\documentclass{emulateapj}

\usepackage{float}
\usepackage{amsmath}
\usepackage{epsfig,floatflt}
\usepackage{subfigure}
\usepackage{rotating}

\begin{document}

\title{Multipole vector anomalies in the
  first-year \emph{WMAP} data: a cut-sky analysis}

\author{P. Bielewicz} 
\affil{Institute of Theoretical Physics, Warsaw University, ul.\ Ho{\.z}a 69, 00-681
Warsaw, Poland}
\email{Pawel.Bielewicz@fuw.edu.pl}

\author{H.\ K.\ Eriksen\altaffilmark{1,2,3}} \affil{Institute of
Theoretical Astrophysics, University of Oslo, P.O.\ Box 1029 Blindern,
\\ N-0315 Oslo, Norway}
\altaffiltext{1}{Also at Centre of Mathematics for Applications,
University of Oslo, P.O.\ Box 1053 Blindern, N-0316 Oslo}
\altaffiltext{2}{Also at Jet Propulsion Laboratory, M/S 169/327, 4800
  Oak Grove Drive, Pasadena CA 91109} 
\altaffiltext{3}{Also at California Institute of Technology, Pasadena, CA
91125}
\email{h.k.k.eriksen@astro.uio.no}

\author{A.\ J.\ Banday}
\affil{Max-Planck-Institut f\"ur Astrophysik, Karl-Schwarzschild-Str.\
1, Postfach 1317,\\D-85741 Garching bei M\"unchen, Germany} 
\email{banday@MPA-Garching.MPG.DE}

\author{K.\ M.\ G\'orski\altaffilmark{3}} 
\affil{Jet Propulsion Laboratory, M/S 169/327, 4800 Oak Grove Drive, 
Pasadena CA 91109\\ Warsaw University Observatory, Aleje Ujazdowskie
4, 00-478 Warszawa, Poland}

\email{Krzysztof.M.Gorski@jpl.nasa.gov}

\and

\author{P.\ B.\ Lilje\altaffilmark{1}} \affil{Institute of Theoretical
Astrophysics, University of Oslo, P.O.\ Box 1029 Blindern, \\N-0315
Oslo, Norway}

\email{per.lilje@astro.uio.no}


\begin{abstract}
We apply the recently defined multipole vector framework to the
frequency-specific first-year \emph{WMAP} sky maps, estimating the
low-$\ell$ multipole coefficients from the high-latitude sky by means
of a power equalization filter. While most previous analyses of this
type have considered only heavily processed (and
foreground-contaminated) full-sky maps, the present approach allows
for greater control of residual foregrounds, and therefore potentially
also for cosmologically important conclusions. The low-$\ell$
spherical harmonics coefficients and corresponding multipole vectors
are tabulated for easy reference.

Using this formalism, we re-assess a set of earlier claims of both
cosmological and non-cosmological low-$\ell$ correlations based on
multipole vectors. First, we show that the apparent $\ell=3$ and 8
correlation claimed by \cite{copi:2004} is present only in the heavily
processed map produced by \cite{tegmark:2003}, and must therefore be
considered an artifact of that map. Second, the well-known
quadrupole-octopole correlation is confirmed at the 99\% significance
level, and shown to be robust with respect to frequency and sky
cut. Previous claims are thus supported by our analysis. Finally, the
low-$\ell$ alignment with respect to the ecliptic claimed by
\cite{schwarz:2004} is nominally confirmed in this analysis, but also
shown to be very dependent on severe \emph{a-posteriori}
choices. Indeed, we show that given the peculiar quadrupole-octopole
arrangement, finding such a strong alignment with the ecliptic is not
unusual.
\end{abstract}

\keywords{cosmic microwave background --- cosmology: observations --- 
methods: numerical}

\maketitle

\section{Introduction}

Since the first-year \emph{Wilkinson Microwave Anisotropy Probe}
(\emph{WMAP}) data release \citep{bennett:2003a}, a great deal of
effort has been spent on analyzing the higher-order statistical
properties of the sky maps. This effort has resulted in several
reports of both non-Gaussianity and statistical anisotropy \citep{de
Oliveira-Costa:2004, eriksen:2004a, eriksen:2004b, eriksen:2005,
hansen:2004a, hansen:2004b, jaffe:2005, larson:2004, vielva:2004},
established by means of many qualitatively different methods. Since
such findings would contradict the currently popular inflationary
based cosmological paradigm, it is of great importance to determine
both their origin and significance.

To aid this work, several new methods have been devised. In
particular, one method was pioneered by Copi, Huterer \& Starkman
(2004), who re-discovered a particular decomposition of a given
multipole into a geometrically more meaningful set of objects, the
so-called (Maxwell's) multipole vectors. Whereas the standard
spherical harmonics expansion is coordinate dependent, these objects
are rotationally invariant, providing a somewhat more intuitive
interpretation of the object. Specifically, the multipole vector set
corresponding to a multipole of order $\ell$ consists of $\ell$ unit
vectors and one overall magnitude.

Since the first paper by \citet{copi:2004}, several other groups have
advanced the method significantly. Choosing a mathematically more
stringent approach, \citet{katz:2004} and \citet{weeks:2004} both
proved uniqueness of the multipole vector decomposition and
established efficient methods for computing it. \citet{land:2005}
focused on the importance of distinguishing between non-Gaussianity
and anisotropy, and introduced the notion of a multipole
frame. Finally, a mathematically elegant approach was taken by
\citet{slosar:2004}, who used a Markov Chain Monte Carlo algorithm to
map out the complete probability distribution of the low-$\ell$
components, and subsequently used these results to study multipole
vector anomalies.

All groups applied their methods to the first-year \emph{WMAP} data
with various results. However, by far most of the effort was spent on
analyzing a small set of heavily processed full-sky maps (the
\emph{WMAP} Internal Linear Combination map [WILC] -- Bennett et al.\
2003a; the Lagrange Internal Linear Combination map [LILC] -- Eriksen
et al. 2004c; the Tegmark et al.\ cleaned map [TOH] -- Tegmark, de
Oliveira-Costa \& Hamilton 2003) which are known to have serious
problems with residual foregrounds \citep{eriksen:2004c}. In fact,
with the exception of the TOH map, the creators of these maps
explicitly warn against using them for cosmological analysis. 

A notable exception among the analyses quoted above is that of
\citet{slosar:2004}. Their approach is statistically sound, in that it
supports partial sky analysis and proper foreground marginalization,
but it is also computationally demanding. Its application is therefore
somewhat limited. Further, their particular choices and treatment of
data make a direct comparison between their results and the ones
presented by other groups somewhat unclear.

It should also be noted that \citet{land:2005} did analyze the proper
\emph{WMAP} maps as well, by applying a sky cut to the data
directly. However, they did not attempt to reconstruct the full-sky
multipole coefficients, and their analysis therefore suffers from
multipole mode coupling and increased error bars.

The popularity of the full-sky maps listed above comes from the fact
that they appear free of foregrounds by visual inspection. The
multipole coefficients may therefore formally be estimated by a
straightforward spherical harmonics expansion, without reference to
any sky cut and subsequent mode decoupling. Nevertheless, even though
it may be difficult to see the foreground residuals by eye, they are
certainly present, and an analysis that fully relies on these maps
will necessarily be cosmologically dubious. In this paper we address
this issue by combining a previously introduced power equalization
filter method for estimating the full-sky spherical harmonics
coefficients from partial sky data with the ordinary multipole vector
method. This allows us to analyze the data frequency by frequency and
region by region. In other words, the multipole vector method may
finally be used for cosmological studies.

The latter analysis takes a very conservative approach to foreground
uncertainties, and, while statistically very robust, the results are
not necessarily directly comparable to the ones obtained by other
groups, primarily due to choice and treatment of the involved data. 

The paper is organized as follows: In Section \ref{sec:methods} we
briefly review the methods used for both estimating the full-sky
harmonics coefficients from cut-sky data, and for computing the
multipole vector decomposition from these. Next, in Section
\ref{sec:data} we describe the data and simulations used in the
analysis. In Section \ref{sec:foregrounds} we study the efficiency of
the power equalization filter method for reconstructing the multipole
components, and compare it with the full-sky cleaning methods. Then we
apply our methods to the first-year \emph{WMAP} data in Section
\ref{sec:results}, seeking to reproduce earlier claims found in the
literature. Concluding remarks are made in Section
\ref{sec:conclusions}. For easy reference, we also tabulate the
low-$\ell$ multipole coefficients for the three cosmologically
interesting \emph{WMAP} Q-, V- and W-bands in Appendix
\ref{sec:lowl_coeff}.

\section{Methods and statistics}
\label{sec:methods}

The following subsections briefly review the methods used in this
paper. We refer the interested reader to the original papers for full
details; Bielewicz, G\'orski \& Banday (2004) for partial sky analysis
by power equalization (PE) filtering, and \citet{copi:2004} for
multipole vector decomposition.

\subsection{Partial sky analysis by PE filtering}

For a given analysis of CMB data to be cosmologically interesting, great
care must be taken to exclude non-cosmological foregrounds. In the
future it may be possible to perform component separation efficiently,
but at present, the only reliable approach is to apply a sky cut and
exclude contaminated pixels from the analysis. 

While the effect of this operation is transparent in pixel space, it
is more complicated in spherical harmonics space, as the spherical
harmonics are no longer orthogonal on a cut sky. In order to estimate
the full-sky harmonics decomposition from partial sky data, one must
therefore decouple the coefficients taking into account the coupling
matrix. PE filtering as described by \citet{bielewicz:2004} is one
method for doing so. 

The first step in this approach is to introduce a new basis set of
functions, $\psi$, that is orthogonal on the cut sky
\citep{gorski:1994}. In this new basis, the vector\footnote{Mapping
from an index pair $(\ell,m)$ into a single index $i$ is given by
\mbox{$i=\ell^2+\ell +m+1$}} of decomposition coefficients,
$\mathbf{c}$, is related to the vector of decomposition coefficients
of the true signal full sky map $\mathbf{a}$ through the relation
\begin{equation} \label{ca_relation} 
\mathbf{c}=\mathbf{L}^T \cdot \mathbf{a}+\mathbf{n}_\psi \ ,
\end{equation}
where $\mathbf{L}$ is the matrix derived by the Cholesky decomposition
of the coupling matrix,
\begin{align}
 \mathbf{K} &= \mathbf{L}\mathbf{L}^{\textrm{T}},\\
K_{i(\ell,m),j(\ell',m')} &= \int_{\rm{cut\ sky}} Y_{\ell m}^\ast
(\hat{\mathbf{n}}) Y_{\ell' m'}^{}(\hat{\mathbf{n}}) d
\Omega_{\hat{\mathbf{n}}},
\end{align}
and $\mathbf{n}_\psi$ is the vector of
noise coefficients in the $\psi$ basis.

The sky cut causes the coupling matrix $\mathbf{K}$ to be singular, as
there is no information in the data about the spherical harmonic modes
that lies fully within the sky cut. Hence it is impossible to
reconstruct all modes from the $\mathbf{c}$. However, for low-order
multipoles, small sky cuts and a high signal-to-noise ratio, it is a
good approximation to simply truncate the vectors and coupling matrix
at some multipole $\ell_{\rm{max}}$, and then reconstruct the
multipole coefficients up to multipole $\ell_{\rm{rec}}$ by filtering
of the data vector $\mathbf{c}_\mathcal{L}$,
\begin{equation}
\hat{\mathbf{a}}_\mathcal{L}=\mathbf{F} \cdot \mathbf{c}_\mathcal{L}.
\end{equation}
Here the subscript $\mathcal{L}$ denotes the range of indices
\mbox{$i=1,\dots,(\ell_{\rm{rec}}+1)^2$}, and the filter $\mathbf{F}$
may be chosen to such that the solution $\hat{\mathbf{a}}_\mathcal{L}$
satisfy a desired set of conditions. In this paper we will consider
the so-called power equalization (PE) filter defined by 
\begin{equation} \label{pe_cond} 
\left<
\hat{\mathbf{a}}_\mathcal{L} \cdot \hat{\mathbf{a}}^T_\mathcal{L}
\right> = \left<\mathbf{a}_\mathcal{L} \cdot \mathbf{a}^T_\mathcal{L}
\right>. 
\end{equation}

To construct the filter for the \emph{WMAP} data we make the usual
assumption that both the CMB and noise components are Gaussian
stochastic variables. Further, the CMB field is assumed to be
isotropic, and the variance of each mode therefore only depends on
$\ell$, $C_\ell=\left<a^2_{\ell m} \right>$. In this paper, we choose
the best-fit (to CMB data alone) \emph{WMAP} power spectrum with a
running spectral index\footnote{Available at
http://lambda.gsfc.nasa.gov}
\citep{bennett:2003a,spergel:2003,hinshaw:2003} as our reference
spectrum. The rms noise level in pixel $p$ is given by
$\sigma(p)_{\rm{noise}}=\sigma_0/\sqrt{N_{\rm{obs}}(p)}$, where
$N_{\rm{obs}}(p)$ is the number of observations per pixel, and
$\sigma_0$ is the rms noise per observation.

Dependence on the assumed power spectrum might seem to be a
disadvantage of this filtering method. However, it was shown by
\citet{bielewicz:2004} that the multipole estimation does not depend
significantly on the assumed power spectrum in the case of the
first-year \emph{WMAP} data. The same applies to the choice of
$\ell_{\rm{rec}}$ and $\ell_{\rm{max}}$. We have chosen
$\ell_{\rm{rec}}=10$ and $\ell_{\rm{max}}=30$, but the multipole
coefficients do not show strong dependence on these parameters.

\subsection{Multipole vector decomposition}

The multipole vector formalism was introduced to CMB analysis by
\citet{copi:2004}, who showed that a multipole moment can be
represented in terms of $\ell$ unit vectors and an overall
magnitude. As later pointed out by \citet{weeks:2004}, the formalism
was in fact first discovered by \citet{maxwell:1891}. Maxwell showed
that for a real function $f_\ell (x,y,z)$ that is an eigenfunction of
the Laplacian on the unit sphere with eigenvalue $-\ell(\ell+1)$
(i.e., spherical harmonic function $Y_{\ell m}$) there exists $\ell$
unit vectors $\mathbf{v}_1,\dots,\mathbf{v}_\ell$ such that:
\begin{equation} 
f_\ell (x,y,z)= A^{(\ell)}\nabla_{\mathbf{v}_1} \dots
\nabla_{\mathbf{v}_\ell}\frac{1}{r} \ ,
\end{equation}
where $\nabla_{\mathbf{v}_i} = \mathbf{v}_i \cdot \nabla$ is the
directional derivative operator and $r=\sqrt{x^2+y^2+x^2}$. A more
useful form of this representation was given by \citet{dennis:2004},
\begin{equation}
f_\ell (\mathbf{r}) = A^{(\ell)} (\mathbf{v}_1 \cdot \mathbf{r}) \dots
(\mathbf{v}_\ell \cdot \mathbf{r}) + Q.
\end{equation}
Here $A^{(\ell)}$ is an overall magnitude, and $Q$ is term fully
defined by the components of $\mathbf{v}_i$ that include components of
angular momenta $\ell-2 \ , \ell-4\ ,\dots$. (This term is needed to
take into account the fact that the product of $\ell$ vectors contains
terms with angular momenta $\ell-2\ , \ell-4\ ,\dots$.) Returning to
the usual language of CMB analysis, each multipole $T_\ell$ may
therefore be uniquely expressed by $\ell$ multipole vectors
$\mathbf{v}^{(\ell,1)}, \ldots, \mathbf{v}^{(\ell,\ell)}$ and a magnitude
$A^{(\ell)}$. In the notation of \citet{copi:2004}, this reads
\begin{equation} 
T_\ell(\hat{\mathbf{e}})\equiv \sum_{m=-\ell}^\ell a_{\ell m} Y_{\ell
m}(\hat{\mathbf{e}}) = A^{(\ell)} (\mathbf{v}^{(\ell,1)} \cdot \hat{\mathbf{e}}) \dots
(\mathbf{v}^{(\ell,\ell)} \cdot \hat{\mathbf{e}}) + Q,
\end{equation}
where $\hat{\mathbf{e}}$ is the radial unit vector in spherical
coordinates. Strictly speaking, the multipole vectors are headless,
thus the sign of each vector can always be absorbed by the scalar
$A^{(\ell)}$. We will use the convention that all vectors point toward
the northern hemisphere.

Algorithms for computing the multipole vectors given a set of $a_{\ell
m}$ coefficients were proposed by \citet{copi:2004} and
\citet{katz:2004}. We have implemented the algorithm of
\citet{copi:2004} in our codes\footnote{The routines of
\citet{copi:2004} are available at
http://www.phys.cwru.edu/projects/mpvectors/}.

\subsection{Multipole vector statistics}

Having computed the multipole vectors, we seek to test a given CMB
data set with respect to either internal correlations between
different multipoles, or external correlations with some given
frame. In order to do so, we follow \citet{copi:2004} and define the
following set of simple statistics.

The first statistic is based on the dot product, which is a natural
measure of vector alignment. Since the multipole vectors are only
defined up to a sign, we choose the absolute value of the dot product
as our statistic. Second, we also consider cross products of the
multipole vectors, $\mathbf{w}^{(\ell,i)} = (\mathbf{v}^{(\ell,j)}
\times \mathbf{v}^{(\ell,k)})$ (where $j \neq k,\ j,k = 1,\dots, \ell$
and $i = 1,\dots,\ell(\ell-1)/2$), in order to test for correlations
between multipole planes. Both normalized and unnormalized cross
products are considered, the latter corresponding to the oriented area
statistic of \cite{copi:2004}. Thus, inspired by \citet{copi:2004} and
\citet{schwarz:2004}, we study the following three statistics for any
two multipoles $\ell_1$ and $\ell_2$ ($\ell_1 \neq \ell_2$):
\begin{itemize}
\item \mbox{$S_{\textrm{vv}}=\sum_{i=1}^{\ell_1} \sum_{j=1}^{\ell_2}
|\mathbf{v}^{(\ell_1,i)}\cdot\mathbf{v}^{(\ell_2,j)}|$},
\item \mbox{$S_{\textrm{vc}}=\sum_{i=1}^{\ell_1}
\sum_{j=1}^{\ell_2(\ell_2-1)/2} |\mathbf{v}^{(\ell_1,i)}\cdot
\mathbf{w}^{(\ell_2,j)}|$},
\item \mbox{$S_{\textrm{cc}}=\sum_{i=1}^{\ell_1(\ell_1-1)/2}
\sum_{j=1}^{\ell_2(\ell_2-1)/2} |\mathbf{w}^{(\ell_1,i)}\cdot
\mathbf{w}^{(\ell_2,j)}|$}.
\end{itemize} 
These are referred to as ``vector-vector'', ``vector-cross'' (or
``cross-vector''), and ``cross-cross'' statistics, respectively.

We use Monte Carlo (MC) simulations to determine the likelihood of
these statistics given the isotropic and Gaussian null-hypothesis (see
Section \ref{sec:data}). As pointed out by \citet{katz:2004} and
\citet{schwarz:2004}, we also note that the multipole vectors of any
given multipole is not internally ordered and neither are the dot
products. Therefore, the sum of dot products is a better statistic
than for instance the $M$ rank-ordered dot products for each
multipoles pair, as defined by \citet{copi:2004}. We will nevertheless
follow the prescription of \citet{copi:2004} in one particular case,
to numerically verify their results. For full details on this
algorithm, we refer the interested reader to the original paper.

\section{Data and simulations}
\label{sec:data}

In this paper, we consider the first-year \emph{WMAP} sky maps
\citep{bennett:2003a} in several forms. Specifically, we analyze both
the template-corrected Q-, V- and W-band frequency maps imposing
various masks (Kp0, Kp2 -- Bennett et al. 2003b; the extended DMR cut,
20+ -- Banday et al. 1997) by means of the PE filter method described
above, and also four heavily processed (and known to be
foreground-contaminated) full-sky maps: the \emph{WMAP} Internal
Linear Combination (WILC) map -- Bennett et al.\ 2003b; the Lagrange
Internal Linear Combination (LILC) map\footnote{Note that the WILC map
is not algorithmically well defined, as the convergence criterion used
for its construction was chosen too liberally, and therefore the
resulting map depends on the initial point for the non-linear
search. Such problems may be avoided by solving the problem using
Lagrange multipliers, and this is done for the LILC map.}-- Eriksen et
al.\ 2004c; the Tegmark, de Oliveira-Costa and Hamilton (TOH) map --
Tegmark et al. 2003; and the latter from which the Doppler quadrupole
term was subtracted -- Schwarz et al.\ 2004. The templates used in the
foreground correction process were those described by
\citet{finkbeiner:1999}, \cite{finkbeiner:2003}, and
\citet{haslam:1982}.

The Doppler term merits some discussion. In principle, this term
should be subtracted from all \emph{WMAP} sky maps prior to
analysis. However, its magnitude is very small indeed, smaller than
both the map making and foreground induced uncertainties (Hinshaw 2005,
private communication), and a result that strongly depends on this
term must therefore necessarily be considered somewhat dubious. We
choose to subtract this term only from the TOH map, in order to assess
its impact.

One goal of this paper is to compare the full-sky ILC method with the
partial-sky PE method. The efficiency of each method is assessed
through simulation, as we plot the true CMB-only statistic value
against the reproduced value after complete processing. The amount of
scatter about the diagonal represents the processing-induced
uncertainty.

A set of 10\,000 ILC simulations were produced using the pipeline
described by \citet{eriksen:2004c}. For the PE method, we used the
simulations described by \citet{bielewicz:2004}.

\section{Algorithmic efficiency and residual foregrounds}
\label{sec:foregrounds}

As pointed out in the introduction, the main short-coming of previous
multipole vector analyses is the fact that they relied on full sky
maps. In this paper we remedy this problem by using the PE filter
method of \citet{bielewicz:2004} to estimate the full-sky harmonics
components from partial sky data. The goal of this section is to study
the relative performance of the PE and the full-sky approaches.

\begin{figure*}

\mbox{\epsfig{file=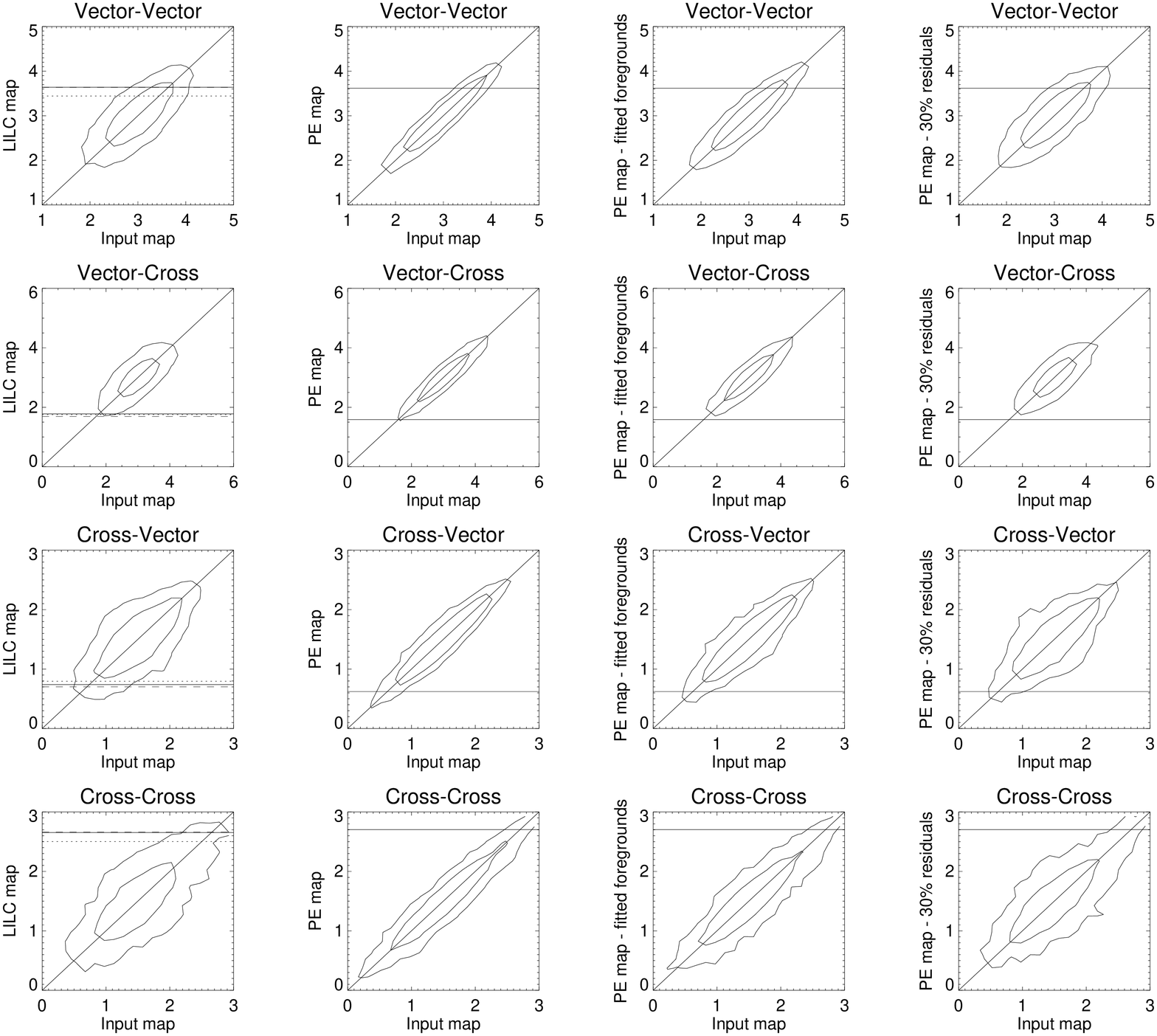,width=\linewidth,clip=}}

\caption{Comparison of the accuracy of the vector-vector, vector-cross
and cross-cross statistics for various methods of the multipoles
estimation.  The abscissae give the true input values of the statistic
for multipoles $\ell_1=2$ and $\ell_2=3$. The ordinates give output
values of the statistic determined after application one of the
method. The contours indicate 68\% and 95\% probability regions as
computed from 10\,000 simulations. The horizontal lines indicate
values of the statistic for the observed \emph{WMAP} data. In the left
column, the lines show the LILC results (solid line), the WILC results
(dotted line), and the TOH results (dashed line). In the three
remaining columns, the lines show the PE-filter results as obtained
from the V-band \emph{WMAP} sky maps combined with the Kp2 sky cut.}
\label{fig:dot_prod}
\end{figure*}

To do so, we apply the above formalism first to 10\,000 LILC
simulations \citep{eriksen:2004c}, and then to 10\,000 PE simulations
\citep{bielewicz:2004}. For each single simulation, we also compute
the same statistic from the pure CMB input full-sky map, and make a
scatter plot of the reconstructed value against the known input
value. For a hypothetical method that is able to reconstruct the input
CMB field perfectly, all points would obviously lie on the diagonal in
such a plot.

Since the PE method supports partial sky analysis, we use the
template-corrected \emph{WMAP} sky maps
for our analysis
in the next sections. This ensures that residual foregrounds are kept
at a minimal level \citep{bennett:2003b}, and the PE simulations are
therefore made without including foregrounds. Nevertheless, the real
template-corrected maps are certainly not \emph{free} of residuals,
and the direct comparison between the LILC and the PE simulations may
therefore be considered somewhat unfair.

To quantify the effect of such residuals, we analyze two additional
sets of simulations. For the first set, we add the templates with the
amplitudes given by \citet{bennett:2003b} (assuming fixed free-free
and synchrotron spectral indices), and then fit for the amplitudes
using the approach described by \citet{gorski:1996}. We then estimate
the low-$\ell$ $a_{\ell m}$'s with the PE filtering method. For the
second set, we add the templates with 30\% of the amplitudes to each
simulation, and re-analyze the simulations. Since we do not attempt to
correct for the foregrounds at all in this case, the latter set of
simulations grossly over-estimates the residual foregrounds present in
the template-corrected maps.

The results from these exercises are shown in Figure
\ref{fig:dot_prod}. Left to right columns show 1) the LILC results, 2)
the clean PE results, 3) the moderately contaminated PE results, and 
4) the heavily contaminated PE results, respectively. Rows show the
four different statistics for multipoles $\ell_1=2$  and $\ell_2=3$ based on 
normalized cross-products.

\begin{deluxetable}{lcccc}
\tablewidth{0pt}
\tablecaption{Signal reconstruction efficiency\label{tab:foregr_corr_stdev}} 
\tablecomments{The correlation coefficients and standard deviation of the
  $S$ statistic for the LILC and PE methods. The row marked PE (1) shows
  results for the PE method applied to clean V-band input maps with Kp2 sky
  coverage, the row PE (2) shows results for input maps with
  subtracted foreground templates with fitted coefficients, and the row 
  PE (3) shows results for the input maps with added
  foreground templates with 30\% of the fit coefficients given by
  \citet{bennett:2003b}. (See text for details.)}
\tablecolumns{5}
\tablehead{ & Vec-Vec & Vec-Cross & Cross-Vec & Cross-Cross}
\startdata

\cutinhead{Pearson's linear correlation coefficient}\\
LILC   &  0.750 & 0.746  & 0.712 & 0.698 \\
PE (1) &  0.962 & 0.953  & 0.955 & 0.943 \\
PE (2) &  0.911 & 0.904  & 0.889 & 0.877 \\
PE (3) &  0.811 & 0.807  & 0.759 & 0.754 \\ 

\cutinhead{Standard deviation}\\
LILC   &  0.325 & 0.411 & 0.277 & 0.355 \\
PE (1) &  0.126 & 0.178 & 0.109 & 0.154 \\
PE (2) &  0.193 & 0.253 & 0.171 & 0.226 \\
PE (3) &  0.281 & 0.358 & 0.253 & 0.321 

\enddata

\end{deluxetable}

Clearly, the PE filter approach is superior to the ILC
approach. Even in the unrealistic case of 30\% residual foregrounds,
the scatter is smaller for the PE filter than for the ILC
method. Realistically, the PE filter performs somewhat worse than the
second column, but slightly better than the third.

We now quantify the scatter observed in each panel both by Pearson's
linear correlation coefficient, and by the standard deviation as
measured orthogonal to the diagonal in each plot. The results from
these computations are summarized in Table
\ref{tab:foregr_corr_stdev}. The visual impression from Figure
\ref{fig:dot_prod} is confirmed by these numbers: the PE filtering
method clearly outperforms the ILC method even in the presence of
unrealistically strong residuals.

Finally, we take the opportunity to once again emphasize that the
full-sky ILC map (and variations thereof) should not be used for
cosmological purposes if it is at all possible to avoid. Such maps are
highly contaminated by residual foregrounds which are likely to have a
significant impact on any moderately sensitive statistic. This is
particularly true on large angular scales, such as the ones discussed
in this and related papers. The massive scatter seen in the left
column of Figure \ref{fig:dot_prod} should serve as a clear indication
of this fact.

\section{Analysis of first-year \emph{WMAP} data}
\label{sec:results}

We now apply our methods to the first-year \emph{WMAP} data, focusing
on three specific claims found in the literature. First, using the
multipole vector approach \citet{copi:2004} found some peculiar
correlations in the $\ell=3,\ldots,8$ range in the TOH map. These
correlations manifested themselves in terms of a number of significant
values of the (normalized and unnormalized) $S_{\textrm{cc}}$
statistic. Here we seek to reproduce these results in the official
template-corrected \emph{WMAP} maps\footnote{Unless explicitly stated
otherwise, all PE results in this section refer to the V-band data
alone, which are the cleanest of the three \emph{WMAP} bands. Noise is
not an issue on the scales of interest, and therefore we do not co-add
the data.}  using the PE filter method.

Second, numerous authors have reported a strong alignment between the
quadrupole and the octopole moments (e.g., de Oliveira-Costa et al.\
2004; Copi et al.\ 2004; Katz \& Weeks 2004) using various methods,
for instance multipole vector alignments. In Section
\ref{sec:quad_oct} we confirm these findings with our improved method.

Finally, a highly surprising claim was made by \citet{schwarz:2004},
who found a nominally strong alignment of the quadrupole and octopole
planes with the ecliptic, and even with the vernal equinox. If
confirmed real, this finding would suggest that the low-$\ell$
anisotropy pattern seen in the \emph{WMAP} data could be of a solar system
origin. This claim is considered in some detail in Section
\ref{sec:ecliptic}.

We note that we have also analyzed our own template-corrected maps
(using the method of G\'{o}rski et al.\ 1996), and found very similar
results as those presented here. Details in the template correction
process are therefore not likely to have a major impact on our
results, assuming that the templates do indeed trace the real
foregrounds satisfactory.

\subsection{Internal correlations among the $\ell=2,\ldots, 8$ multipoles}

The first analysis of the \emph{WMAP} data based on multipole vectors
was performed by \citet{copi:2004}. The main conclusion from this work
was a claim of correlations among the low-$\ell$ multipoles in
contradiction with the currently preferred Gaussian and isotropic
cosmological model. This claim was based on two observations. First,
the \emph{normalized} cross-cross multipole vector statistic indicated
a very strong correlation between the $\ell=3$ and 8 modes, and
second, the \emph{unnormalized} cross-cross statistic revealed five or
eight (depending on likelihood threshold) moderately strong
correlations within the $\ell=2$ to 8 modes.

While the nominal significance of their detection was reasonably high
(roughly at the 99\% level), two problems could be identified with the
analysis. First, as subsequently pointed out by several authors
\citep{katz:2004,schwarz:2004}, their statistics were based on a
rather elaborate rank ordering scheme with a somewhat unclear
interpretation. It is not clear how robust this method is. Second, and
more importantly, only the foreground-contaminated WILC and TOH maps
were considered in the analysis.

In this paper, we first repeat the original analysis of
\citet{copi:2004} based on rank ordering to see if the results (at
least nominally) hold when applied to the template-corrected \emph{WMAP}
maps. However, we also analyze the same multipole pairs with the $S$
statistics, in order to study the statistical robustness of the
detection. For specifics on the procedure, we refer to
\citet{copi:2004}; our main target in this paper is robustness with
respect to foregrounds, not algorithmic consistency.

The results from the rank ordering analysis are summarized in Table
\ref{tab:prob_rank}. Columns 2 through 5 are to be compared with
column 8 of Table 1 of \citet{copi:2004}, while columns 6 through 9
are to be compared with columns 10 and 12. While the agreement between
our TOH-DQ numbers and their numbers is not perfect, it is quite good
overall. Tracking down the cause of the small differences would
require having access to both codes; minor details such as the bin
size used for estimating the likelihoods do have an impact, in
particular on probabilities not in the tails of the
distributions. However, we note that we observe perfect agreement with
previously published $S$ statistic results \citep{schwarz:2004,
katz:2004, weeks:2004} for all published cases, so all codes appear to
be working as expected.

\begin{deluxetable*}{ccccccccccc}
\tablewidth{0pt}
\tablecaption{Low-$\ell$ multipole correlations by rank ordering\label{tab:prob_rank}} 
\tablecomments{Probabilities of obtaining a lower cross-cross statistic
  value than that observed in the first-year \emph{WMAP} data,
  measured relative to MC simulations by means of \emph{rank ordering}
  \citep{copi:2004}. The statistics were computed for the LILC, WILC,
  TOH-DQ, and PE filtered V-band \emph{WMAP} maps.
}
\tablecolumns{11}
\tablehead{&&\multicolumn{4}{c}{Normalized Cross-Cross} &$\quad$&
\multicolumn{4}{c}{Unnormalized Cross-Cross} \\
$(\ell_1, \ell_2)$ && LILC & WILC & TOH-DQ & PE && LILC & WILC & TOH-DQ & PE}
\startdata

$(2, 3)$ &&  \phn1.73\% & \phn5.04\% & \phn2.19\% & \phn1.45\% & & \phn1.75\% & \phn1.38\% & \phn0.11\% & \phn3.62\% \\ 
$(2, 4)$ &&  28.16\% & 55.78\% & 58.73\% & 43.21\% & & 25.28\% & 83.86\% & 89.44\% & 57.50\% \\	
$(2, 5)$ &&  79.97\% & 77.17\% & 23.15\% & 34.36\% & & 99.22\% & 91.30\% & 66.77\% & 76.92\% \\	
$(2, 6)$ &&  81.53\% & 88.36\% & 87.40\% & 37.27\% & & 64.73\% & 96.36\% & 78.55\% & 19.57\% \\	
$(2, 7)$ &&  91.08\% & 99.56\% & 93.20\% & 79.47\% & & 91.54\% & 94.03\% & 86.24\% & 68.07\% \\	
$(2, 8)$ &&  82.55\% & 60.75\% & 89.81\% & 47.06\% & & 67.23\% & 94.84\% & 68.29\% & 28.47\% \\	
$(3, 4)$ &&  \phn8.28\% & \phn6.56\% & 17.02\% & 69.95\% & & 14.97\% & 12.61\% & 30.85\% & 70.68\% \\	
$(3, 5)$ &&  45.95\% & 70.13\% & 37.40\% & 30.05\% & & 58.86\% & 74.46\% & 65.30\% & 64.63\% \\	
$(3, 6)$ &&  57.59\% & 46.74\% & 73.13\% & 32.75\% & & 75.31\% & 55.18\% & 92.67\% & 70.25\% \\	
$(3, 7)$ &&  59.28\% & 70.55\% & 53.90\% & 89.17\% & & 86.54\% & 84.12\% & 74.25\% & 75.14\% \\	
$(3, 8)$ &&  88.47\% & 86.21\% & 99.99\% & 51.26\% & & 97.75\% & 98.33\% & 99.75\% & 52.92\% \\	
$(4, 5)$ &&  45.67\% & 48.51\% & 53.44\% & 35.19\% & & 70.24\% & 68.26\% & 53.26\% & 37.53\% \\	
$(4, 6)$ &&  97.59\% & 86.04\% & 74.28\% & 56.19\% & & 87.02\% & 72.28\% & 62.90\% & 40.57\% \\	
$(4, 7)$ &&  58.15\% & 97.56\% & 53.43\% & 82.42\% & & 24.69\% & 32.10\% & 88.82\% & 98.15\% \\	
$(4, 8)$ &&  81.16\% & 88.23\% & 85.67\% & 79.73\% & & 86.53\% & 95.20\% & 99.00\% & 69.17\% \\	
$(5, 6)$ &&  94.28\% & 99.93\% & 66.08\% & 59.03\% & & 95.95\% & 85.30\% & 83.88\% & 24.12\% \\	
$(5, 7)$ &&  99.70\% & 95.69\% & 59.87\% & 84.32\% & & 98.19\% & 98.45\% & 69.04\% & 95.96\% \\	
$(5, 8)$ &&  81.17\% & 78.64\% & 31.27\% & 63.85\% & & 30.90\% & 33.36\% & 24.40\% & 76.85\% \\	
$(6, 7)$ &&  55.27\% & 59.56\% & 20.41\% & \phn8.19\% & & 63.74\% & 77.77\% & 97.42\% & 16.48\% \\	
$(6, 8)$ &&  56.28\% & 80.03\% & 72.51\% & 76.98\% & & 48.28\% & 66.79\% & 62.53\% & 51.73\% \\	
$(7, 8)$ &&  \phn7.99\% & 18.78\% & \phn9.70\% & \phn7.69\% & & 10.26\% & 24.76\% & 25.11\% & \phn5.18\% 	

\enddata

\end{deluxetable*}

As mentioned above, the conclusions of \citet{copi:2004} can be
summarized in terms of two different anomalies. First, a high
significance was observed for the rank-ordered normalized cross-cross
statistic when applied to the $\ell=(3,8)$ pair. This result is
confirmed in Table \ref{tab:prob_rank}, \emph{but only for the TOH
map}. For the other three maps, this particular value is highly
insignificant. Therefore, if this detection does signify a real
feature, it is only present in the foreground-contaminated TOH map,
and not in the more trust-worthy \emph{WMAP} V-band map. It can therefore not
be taken as representative for the first-year \emph{WMAP} data as a whole.

The second anomaly was defined in terms of an unusually large number
of high unnormalized cross-cross values for the TOH map: five out of
21 ranks for the TOH map were found to be larger than 0.9, and eight
out of 21 values were larger than 0.8. Once again, we see that this
conclusion only hold for the TOH map, and not for the PE-filtered
V-band map. Quite on the contrary, the V-band PE filter results appear
completely uniform, and no anomalies can be readily identified. 

\begin{deluxetable*}{cccccc}
\tablewidth{0pt}
\tablecaption{Low-$\ell$ multipole correlations by dot products\label{tab:prob_methods_unnorm}} 
\tablecomments{Probabilities of finding a value of the $S_{\textrm{cc}}$
  statistic lower than the observed \emph{WMAP} values, estimated from
  ensembles of 10\,000 MC simulations.  The cross-products are all
  unnormalized (corresponding to the ``oriented area'' statistic of
  Schwarz et al.\ 2004), except for the last column. The PE filter
  results were obtained from the V-band alone, imposing the Kp2 mask.}
\tablecolumns{6}
\tablehead{$(\ell_1, \ell_2)$ & LILC & WILC & TOH-DQ & PE & PE (norm) }

\startdata
$(2, 3)$ & 97.16\% &  99.29\% &  99.87\% &  95.43\%  & 98.82\%  \\
$(2, 4)$ & 18.06\% &  36.71\% &  34.72\% &  26.77\%  & 41.30\%  \\
$(2, 5)$ & 31.94\% &  64.20\% &  51.29\% &  18.24\%  & 32.34\%  \\
$(2, 6)$ & 22.30\% &  63.83\% &  62.02\% &   \phn8.64\%  & 15.88\%  \\
$(2, 7)$ & 27.44\% &  49.98\% &  79.38\% &  30.72\%  & 82.54\%  \\
$(2, 8)$ & 23.71\% &  45.93\% &  76.57\% &  13.70\%  & 22.63\%  \\
$(3, 4)$ &  \phn8.19\% &   \phn6.31\% &  11.74\% &  31.43\%  & 28.21\%  \\
$(3, 5)$ & 66.52\% &  55.50\% &  55.61\% &  49.25\%  & 57.42\%  \\
$(3, 6)$ & 63.03\% &  67.86\% &  71.35\% &  37.73\%  & 78.77\%  \\
$(3, 7)$ & 57.68\% &  53.91\% &  66.51\% &  41.84\%  & 70.52\%  \\
$(3, 8)$ & 51.58\% &  38.07\% &  47.64\% &  22.06\%  &  \phn8.36\%  \\
$(4, 5)$ & 66.54\% &  68.05\% &  75.88\% &  92.05\%  & 67.29\%  \\
$(4, 6)$ & 42.59\% &  33.51\% &  57.19\% &  32.64\%  & 51.10\%  \\
$(4, 7)$ & 21.56\% &  23.02\% &  46.10\% &  45.83\%  & 46.38\%  \\
$(4, 8)$ & 44.77\% &  48.59\% &  57.79\% &  71.25\%  & 45.38\%  \\
$(5, 6)$ & 70.94\% &  82.05\% &  79.16\% &  24.32\%  & 31.81\%  \\ 
$(5, 7)$ & 64.26\% &  73.48\% &  77.59\% &  42.08\%  & 46.88\%  \\
$(5, 8)$ & 97.68\% &  97.32\% &  97.86\% &  89.23\%  & 54.70\%  \\
$(6, 7)$ & 64.16\% &  77.42\% &  64.94\% &   \phn6.83\%  & 92.13\%  \\
$(6, 8)$ & 87.23\% &  90.25\% &  73.87\% &  14.50\%  & 83.13\%  \\
$(7, 8)$ &  \phn8.67\% &  12.84\% &  30.35\% &   \phn2.74\%  &  \phn8.11\%  
\enddata

\end{deluxetable*}

Even though the above results appear to refute the original claims of
low-$\ell$ correlations, we compute the more robust $S_{\textrm{cc}}$
statistic for the same low-$\ell$ pairs for completeness. The results
from these calculations are shown in Table
\ref{tab:prob_methods_unnorm}. Again, the results appear quite uniform
(with the exception of $\ell=(2,3)$, to which we return in the next
section), and even the $\ell=(3,8)$ pair does not appear anomalous for
any of the maps.

Based on this analysis, we conclude that the results of
\cite{copi:2004} are due to a combination of a poorly defined
statistic and unreliable data. The anomaly does not survive when
subjected to a more careful statistical analysis.

\subsection{Quadrupole and octopole correlations}
\label{sec:quad_oct}

\begin{deluxetable}{lcccc}
\tablewidth{0pt}
\tablecaption{Quadrupole-octopole correlations\label{tab:prob_l2_3}} 
\tablecomments{Probabilities of finding a value of the $S$
  statistic for $(\ell_1,\ell_2)=(2,3)$ lower than that of the observed \emph{WMAP}
  data for various foreground cleaning methods, frequencies and sky
  cuts. The default sky mask for the PE filter method is Kp2. The
  row marked by PE shows results for the PE method applied to the
  template-corrected V-band \emph{WMAP} map (see text for details).
  The cross-products are normalized, following
  \citet{weeks:2004}.}
\tablecolumns{5}
\tablehead{Data & Vec-Vec & Vec-Cross & Cross-Vec & Cross-Cross }

\startdata
\cutinhead{Sensitivity to method}\\
LILC    & 92.04\% & 2.47\% & 2.27\% & 98.57\% \\
WILC    & 83.28\% & 2.24\% & 2.94\% & 97.31\% \\
TOH-DQ  & 90.69\% & 2.03\% & 1.82\% & 98.65\% \\
PE      & 91.36\% & 1.62\% & 1.05\% & 98.82\% \\
\cutinhead{Sensitivity to frequency}\\
Q band & 93.13\% & 0.44\% & 0.45\% & 99.61\% \\
V band & 91.36\% & 1.62\% & 1.05\% & 98.82\% \\
W band & 89.28\% & 1.74\% & 1.30\% & 98.73\% \\
\cutinhead{Sensitivity to sky coverage}\\
Kp2 & 91.36\% & 1.62\%  &  1.05\% & 98.82\% \\
Kp0 & 89.66\% & 2.06\%  & 37.91\% & 89.92\% \\
20+ & 90.43\% & 8.95\%  & 57.92\% & 81.45\% 
\enddata

\end{deluxetable}

A similar type of anomaly was reported by \citet{de
Oliveira-Costa:2004}. They found that the octopole ($\ell=3$ mode) of
the \emph{WMAP} data spans a plane on the sky, and further, that the normal
to this plane is strongly aligned with the quadrupole plane. This
anomaly has since then been extensively studied with different
techniques (e.g., Weeks 2004; Eriksen et al.\ 2004c), and is well
established by now.

The multipole vector framework is particularly suited for this
particular anomaly. Recalling that the
multipole vectors of order $\ell$ contain terms of
orders $\ell, \ell-2, \ell-4, \ldots$, we see that the quadrupole only consists of a 
quadratic term plus a constant, while the octopole consists of a cubic and a linear term. 
Further, it can be shown that the cross-vectors for a given multipole point toward the 
saddle points of the term of order $\ell$ of each multipole (but, unfortunately, not toward
the saddle points of the multipole as a whole.) For the quadrupole, a quite intuitive
interpretation of the two multipole vectors is therefore readily available: their cross-product 
point towards the saddle point. For \emph{WMAP}, a similar statement is very nearly true for even the three
octopole cross-vectors. 

Based on these observations, we can make a 
connection between the multipole vector framework and the approach taken by \citet{de Oliveira-Costa:2004};   
since the \emph{WMAP} octopole is planar, its saddle points are clustered, and all its
cross-products point roughly towards the same point on the
sky. Furthermore, since the quadrupole plane is aligned with the
octopole plane, even this cross-product points towards the same
point on the sky.
This is clearly seen in Figure \ref{fig:maps}, where we plot the positions of the quadrupole and
the three octopole cross-product vectors on the sky, in ecliptic
projection. Clearly, the four vectors are strongly clustered on the
sky, as discussed above. 

To assess the estimator uncertainty (i.e., due to the sky cut) in the
position of each of the low-$\ell$ cross-product vectors, we used the
foreground-free Monte Carlo PE simulations described earlier. For each
simulation, we computed the cross-products from both the input map and
the reconstructed PE-filtered map (for the V-band alone), and computed
the absolute angular distance between the input and output vectors for
all possible pairings, and chose the relative ordering with the
smallest sum of errors. (This is necessary because the multipole
vectors are not internally ordered.) Such computations show that the
mean angular error is about $4^{\circ}$ for the quadrupole
cross-vector, and 6--$9^{\circ}$ for the three octopole vectors. Thus,
the error in each case roughly equals the size of each dot in Figure
\ref{fig:maps}.

Returning to the quadrupole-octopole anomaly, we note that the
previously defined $S$ statistics involving cross-products are well
suited for measuring the degree of alignment for these two modes, due
to the above argument. Following \citet{weeks:2004}, we therefore
adopt the normalized $S_{\textrm{cc}}$ statistic for this particular
analysis, and the corresponding results are tabulated in Table
\ref{tab:prob_l2_3}.

In the top section, results for different foreground cleaning methods
are given. Clearly, the quadrupole-octopole alignment is quite stable
(although not perfectly so) with respect to foreground cleaning
method: the results for the cross-product type statistics are all at
the 98\% confidence level, in good agreement with the 98.7\%
significance obtained using the angular momentum dispersion statistic
of \citet{de Oliveira-Costa:2004}.

In the middle section, we list the same statistic from each of the
three cosmologically interesting \emph{WMAP} frequency bands using the
PE filter. Again, the results are very stable, and this gives us
confidence that the effect is indeed a feature of the CMB field,
rather than caused by residual foregrounds. Further, we also point out
that this particular set of results clearly demonstrates the strength
of the PE filter method; while the other methods only allow for
frequency averaged conclusions, the PE method can provide frequency
specific results, and therefore much greater control over foregrounds.

Finally, in the bottom section of Table \ref{tab:prob_l2_3} we give
the PE filter results for different sky cuts as applied to the V-band
\emph{WMAP} data. (Here we note that the PE estimator uncertainties
for the large 20+ cuts are considerably larger than for the Kp2 and
Kp0 masks, as discussed by Bielewicz et al.\ 2004, and the numbers are
only included here for completeness.)

\subsection{Ecliptic correlations}
\label{sec:ecliptic}

\begin{figure*}

\mbox{\epsfig{file=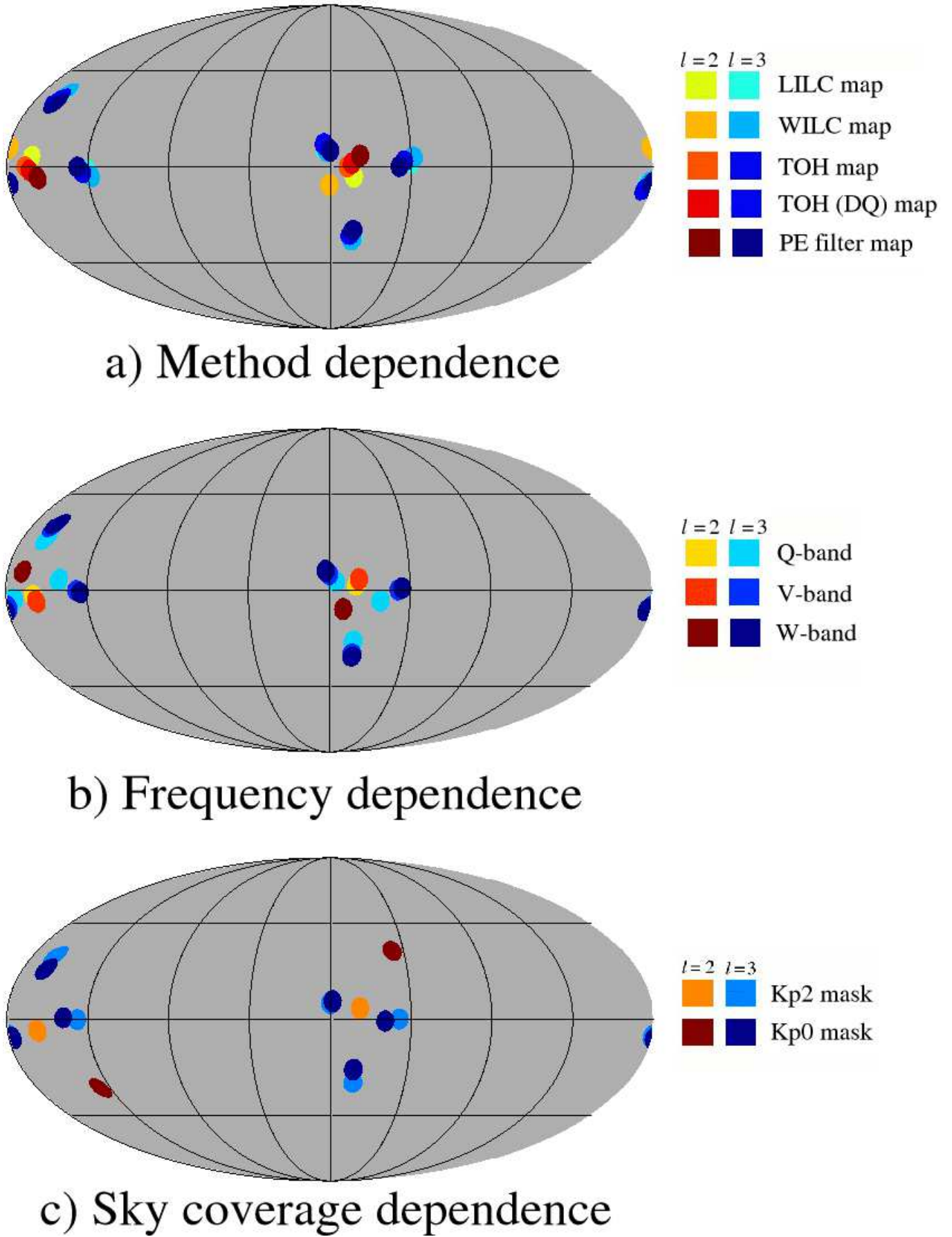,width=\linewidth,clip=}}

\caption{Multipole cross-product vectors on the sky as a function of
  cleaning method (top panel), frequency band (middle panel) and sky
  coverage (bottom panel). Default options for the PE filter results
  are the V-band map and Kp2 sky cut. The maps are shown in the
  ecliptic reference frame with the vernal equinox at the center. The
  radius of each dot is $5^{\circ}$.}
\label{fig:maps}
\end{figure*}

Finally, we consider a set of claims made by \citet{schwarz:2004};
that the low-$\ell$ anisotropy pattern observed by \emph{WMAP} could
have a very local origin, and that there could be yet unknown
microwave sources or sinks within our own solar system. These claims
were based on measuring alignments between the multipole vector
cross-products for $\ell=2$ and 3 and a pre-defined set of fixed
axes. These axes ranged from the somewhat plausible (the
super-galactic and ecliptic) to the highly surprising (the
equinoxes). Their main result was that the four $\ell=2$ and 3
cross-product vectors were nearly orthogonal to the ecliptic
north-south axis, as measured by the dot product. This can visually be
seen in Figure \ref{fig:maps}, as the dots all lie along the equator
in the ecliptic frame, and, indeed, clustered near the vernal
equinox.

We now repeat the calculations of \citet{schwarz:2004}, applying the
PE filter method to the Q-, V- and W-band \emph{WMAP} maps, and
imposing the Kp2, Kp0 and 20+ sky cuts. We compute the sum of dot
products between the ecliptic north-south axis and the union of the
$\ell=2$ and 3 cross-product vectors, following \citet{schwarz:2004},
and also for each mode individually, up to $\ell=6$. The results from
these computations are summarized in Table \ref{tab:prob_ecl}.

Again starting with the top section, we see that the numbers (except
for the quadrupole alone) are not very sensitive to the particular
foreground correction method. Further, for the particular combination
in question ($\ell=2$ and 3), the alignment is in fact stronger for
the PE filtered V-band map than for the more contaminated maps
(although this may be somewhat of a coincidence, looking at the
$\ell=2$ and 3 numbers individually). The numbers also do not depend
strongly on frequency or sky cut. As far as these numbers are
concerned, the alignment must therefore be assumed to be of CMB
origin, and not of foreground origin. Taken at face value, these
results therefore appear to confirm the claims made by
\citet{schwarz:2004}.

However, while the nominal significance of the results seems solid, a
much more fundamental objection may be raised against this detection,
namely its strong dependence on \emph{a-posteriori} choices. Two
particular problems may be identified, namely the choice of multipoles
to include, and the choice of external axis.

In the first case, we see in Table \ref{tab:prob_ecl} that the
ecliptic alignment is only significant if one takes into account both
$\ell=2$ and $\ell=3$ simultaneously, and no other
multipoles. Further, the quadrupole-ecliptic alignment alone is only
significant in the TOH map, and quite insignificant in the PE filtered
map. In fact, the numbers for the quadrupole alignment (both for
different methods and for different frequencies) imply that a
significant amount of foreground residuals is present in this mode,
and that its true direction is not well constrained. Results that
strongly depends on this mode cannot be trusted.

\begin{deluxetable}{lcccccc}
\tablewidth{0pt}
\tablecaption{Low-$\ell$-ecliptic correlations\label{tab:prob_ecl}} 
\tablecomments{Probabilities of finding a value of the $S_{\textrm{cc}}$
  statistic lower than the observed \emph{WMAP}
  data for various $\ell$'s, foreground cleaning methods, frequencies and sky
  cuts. Default sky cut for the PE filter method is Kp2. The row
  marked by PE shows
  results for the PE method applied to the template-corrected V-band \emph{WMAP} map (see text for details).
  The cross-products are normalized, following
  \citet{weeks:2004}.
}
\tablecolumns{7}
\tablehead{Data & $\ell=2$ & $\ell=3$ & $\ell=2\!+\!3$ & $\ell=4$ & $\ell=5$ & $\ell=6$}

\startdata
\cutinhead{Sensitivity to method}\\
LILC    &  7.7\% &  3.3\% &  1.0\% &  61.8\% &  20.5\% &  41.6\% \\
WILC    & 13.6\% &  4.0\% &  1.5\% &  67.0\% &  29.5\% &  41.0\% \\
TOH     &  0.0\% &  3.6\% &  0.7\% &  76.8\% &  16.9\% &  42.4\% \\
TOH-DQ  &  2.6\% &  3.6\% &  0.9\% &  76.8\% &  16.9\% &  42.4\% \\
PE      &  9.2\% &  2.3\% &  0.6\% &  71.4\% &  19.8\% &  65.6\% \\
\cutinhead{Sensitivity to frequency}\\
Q band &   3.9\% &  1.8\% &  0.4\% &  44.4\% &   9.9\% &  32.1\% \\
V band &   9.2\% &  2.3\% &  0.6\% &  71.4\% &  19.8\% &  65.6\% \\
W band &  14.6\% &  2.9\% &  1.2\% &  56.7\% &  16.9\% &  57.5\% \\
\cutinhead{Sensitivity to sky coverage}\\
Kp2 &   9.2\% &  2.3\% &   0.6\% & 71.4\%  &  19.8\% &  65.6\% \\
Kp0 &  52.0\% &  1.7\% &   3.4\% & 88.5\%  &  20.9\% &  51.3\% \\
20+ &  58.1\% &  5.0\% &  12.3\% & 34.2\%  &  16.0\% &  53.3\% 
\enddata

\end{deluxetable}

As far as the choice of axis goes, it is important to remember that
the ecliptic axis was identified after looking at the data. It is
therefore very difficult to assess the true significance of the
alignment -- the set of possible choices one could have considered is
indeed large. However, some quantification may be provided by means of
the following arguments.

First, it is important to remember that no known non-cosmological
physical mechanism is able to produce a frequency-independent
signature similar to the one discussed here. A very good
null-hypothesis is therefore that the internal correlations seen in
the CMB pattern are in fact of cosmological origin. Next, as discussed
in the previous section, it is well known by now that \citep{de
Oliveira-Costa:2004}
\begin{enumerate}
\item the octopole moment is somewhat planar, and
\item that the quadrupole plane is strongly aligned with the octopole
  plane.
\end{enumerate}
Again, as described in the previous section, the first point implies
that the three octopole cross-vectors are aligned along some axis, and
the second point implies that the quadrupole cross-vector is aligned
along the same axis. Thus, all four cross-vectors point toward roughly
the same point on the sky. Such arrangements could be established
either by means of cosmological physics (e.g., non-trivial topologies,
cosmic vorticity/shear etc.) or by local physics (e.g., galactic
foregrounds).

The correct question to answer is then, \emph{given} such an
arrangement of the low-$\ell$ multipoles, what is the probability of
finding a stronger alignment with the ecliptic than the observed one?
Or rather, since we presumably would have been equally ``happy'' with
an alignment with the Galactic or super-Galactic reference frames, we
ask, what is the probability of finding a stronger alignment with any
one of the three frames?

To answer this question, we run the following experiment. We take the
set of four observed cross-vectors, and rotate them jointly by an
arbitrary Euler matrix, conserving the relative arrangement but
randomizing the overall orientation and position. This operation is
repeated one million times, each time computing the dot products with
each of the three reference axes, and recording the number of times
any one of these is smaller than the observed ecliptic alignment.

For the Doppler-corrected TOH map we find a stronger alignment in 3\%
of the simulations, and the anomaly can therefore be considered to be
statistically robust. However, for the PE filtered maps we find a
stronger alignment in everywhere from 3 to 39\%, depending on sky cut
and frequency. Once again, the anomaly is therefore considerably
stronger in the TOH map than in the PE filtered maps.

The large variation among the PE filtered maps stems from the fact
that the statistic is highly sensitive to the relative orientation of
all four vectors: a higher significance is found when three of the
four vectors lie on a single great circle, than, for instance, when
the fourth point lies well inside the triangle spanned by the other
three points. Thus, the foreground-sensitive quadrupole vector does
play a significant role in this anomaly, and the particularly strong
quadrupole-ecliptic anti-alignment seen in the TOH map alone is a
strong factor\footnote{Here we also note that although
\citet{schwarz:2004} did in fact consider the quadrupole stability
issue by adding Gaussian noise with rms of $10\,\mu\textrm{K}$ to
$a_{20}$, we believe that this estimate significantly underestimates
the true quadrupole uncertainty in the ILC maps
\citep{eriksen:2004c}.}.

To summarize, from the above experiments it appears that it is not the
external alignment with the ecliptic that is anomalous, but rather the
internal alignments between the quadrupole and octopole: Given such an
arrangement, it is not unlikely to hit upon one of the three most
important reference frames.

A second observation is that the cross-vectors point toward the
ecliptic plane, not the poles. Presumably, an alignment with the poles
would have been even more exciting than an alignment with the plane,
and therefore a two-sided distribution should be considered when
quoting confidence limits. This further reduces the significance of
the anomaly.

In conclusion, it seems unreasonable to us to accept a marginally
significant ($\sim$99\%) effect as physical in light of the numerous
problems connected to it. We believe that it is unnecessary to
introduce the (exceedingly difficult to explain) idea of ecliptic
alignment in addition to the more general quadrupole-octopole
alignment. Of course, local physics may certainly have a role to play
with respect to the latter problem, but Galactic or extra-Galactic
contamination seems like far more plausible candidates than
contamination of solar system origin.

\section{Conclusions}
\label{sec:conclusions}

In this paper, we have revisited a set of claims found in the
literature regarding the low-$\ell$ CMB pattern and multipole
vectors. We have remedied the most serious outstanding problem
connected to these analyses, in that we have used only partial sky
data to estimate the multipole vectors. This allowed us to study the
frequency-specific \emph{WMAP} sky maps individually, while imposing
different sky cuts to study regional dependence. Using these methods,
the multipole vector approach may finally be used for cosmological
analysis.

Three claims were studied in depth. First, \citet{copi:2004} found a
set of strong correlations among the $\ell=2,\ldots,8$ multipoles
using the multipole vector formalism. Unfortunately, they only had
access to two full-sky maps (the WILC and TOH sky maps), which are
known to be contaminated by galactic foregrounds. While we reproduced
their results for these two maps, we also found that the anomaly is
not present in the best available frequency-specific CMB
maps. Therefore, as far as the low-$\ell$ correlations are
statistically significant, they must be considered an artifact of the
TOH and WILC sky maps, and not of the \emph{WMAP} data as a whole.

Second, we revisited the much more established anomaly first reported
by \citet{de Oliveira-Costa:2004}; the strong alignment between the
quadrupole and octopole moments. Our results confirm previous
conclusions: The effect is significant at the 98-99\% confidence
level, and independent of frequency and sky cut. It appears to be
quite robust.

Finally, we also considered the claims made by \citet{schwarz:2004},
that the low-$\ell$ CMB field could be of solar system origin. This
claim was based on the observation that the $\ell=2$ and 3 multipole
cross-product vectors align with the ecliptic north-south axis, and,
indeed, that they point towards the vernal equinox. While the nominal
significance of these results are confirmed in this paper, we also
found that it is not at all unusual to observe such a strong alignment
with one of the three major axes (ecliptic, galactic or
super-galactic), \emph{given} the peculiar internal arrangements of
the quadrupole and octopole. Thus, it is not the ecliptic correlation
\emph{per se} that is anomalous, but rather the quadrupole-octopole
alignment. Whether this latter feature is caused by cosmological or
non-cosmological physics is not yet clear, but solar-system physics
does not appear to provide the most plausible explanation.

\section*{Acknowledgments}

The authors thank Jeff Weeks, Gary Hinshaw, Craig Copi, Dragan
Huterer, Glenn Starkman and Dominik Schwarz for interesting
discussions. PB thanks for encouragement to take up these studies from
M.\ Demia{\'n}ski.  He also thanks Warsaw University Astronomical
Observatory for its hospitality.  HKE thanks Dr.\ Charles R. Lawrence
for his support, and especially for arranging HKE's visit to JPL. He
also thanks the Center for Long Wavelength Astrophysics at JPL for its
hospitality while this work was completed. PB acknowledges financial
support from the Polish State Committe for Scientific Research grant
1-P03D-014-26.  HKE acknowledges financial support from the Research
Council of Norway, including a Ph.\ D. scholarship.  We acknowledge
use of the HEALPix software \citep{gorski:2005} and analysis package
for deriving the results in this paper.  We also acknowledge use of
the Legacy Archive for Microwave Background Data Analysis (LAMBDA).
This work has received support from The Research Council of Norway
(Programme for Supercomputing) through a grant of computing time.
This work was partially performed at the Jet Propulsion Laboratory,
California Institute of Technology, under a contract with the National
Aeronautics and Space Administration.

\appendix

\section[]{The low-$\ell$ PE multipole coefficients of the first-year
  \emph{WMAP} data}
\label{sec:lowl_coeff}

In this Appendix, we tabulate the low-$\ell$ spherical harmonics
coefficients and multipole vectors as computed with the PE filter
method. The methods used in these computations are described by
\citet{bielewicz:2004} for PE filtering, and \citet{copi:2004} or
\citet{weeks:2004} for multipole vector estimation.

\begin{deluxetable*}{cccc}
\tablewidth{0pt}
\tablecaption{Low-$\ell$ spherical harmonics coefficients\label{tab:lowl_coeff}} 
\tablecomments{Full-sky low-$\ell$ spherical harmonics coefficients
  reconstructed from the high-latitude template-corrected \emph{WMAP} data by means of the PE filter
  method of \citet{bielewicz:2004}.
}
\tablecolumns{4}
\tablehead{ Multipole  & Q-band & V-band & W-band \\
  $(\ell,m)$ & $(\mu\textrm{K})$ & $(\mu\textrm{K})$ & $(\mu\textrm{K})$  \\
}
\startdata

(2,0) & ( 11.53 +  0.00$i$ ) & ( 15.57 +   0.00$i$ ) & ( 10.14 +   0.00$i$ )  \\
(2,1) & ( -5.54 +  3.09$i$ ) & ( -5.37 +   2.37$i$ ) & ( -4.94 +   2.98$i$ ) \\
\vspace*{2mm}(2,2) & ( -9.52 - 15.89$i$ ) & ( -12.31 - 17.78$i$ ) & ( -13.46 - 18.54$i$ )  \\

(3,0) & ( \phn-6.92 \phm{-}+   \phn0.00$i$ ) & ( -5.70 +   0.00$i$ ) & ( -5.41 +   0.00$i$ )  \\
(3,1) & ( \phn-4.53  \phm{+}- \phn1.53$i$ ) & ( -9.06  - 0.18$i$ ) & ( -9.50 +   0.78$i$ )  \\
(3,2) & ( \phm{-}23.27 + \phn0.12$i$ ) & ( 21.95 +   0.98$i$ ) & ( 22.04 +   0.74$i$ )  \\
\vspace*{2mm}(3,3) & ( -20.57 +  28.58$i$ ) & ( -15.68 +  29.61$i$ ) & ( -14.64 +  29.40$i$ )  \\

(4,0) & ( 19.60 +   0.00$i$ ) & ( 15.57 +   0.00$i$ ) & ( 21.31 +   0.00$i$ ) \\
(4,1) & ( -4.83 +   9.92$i$ ) & ( -7.15 +   9.21$i$ ) & ( -7.71 +   8.38$i$ ) \\
(4,2) & (  7.55 +   6.76$i$ ) & (  9.31 +   8.05$i$ ) & (  9.35 +   8.32$i$ )  \\
(4,3) & (  2.68  - 21.94$i$ ) & (  5.21  - 21.75$i$ ) & (  6.24  - 20.75$i$ )  \\
\vspace*{2mm}(4,4) & ( 10.82   - 5.30$i$ ) & (  5.89   - 7.75$i$ ) & (  4.70   - 9.36$i$ )  \\

(5,0) & ( 16.05 +   0.00$i$ ) & ( 15.46 +   0.00$i$ ) & ( 14.35 +   0.00$i$ )  \\
(5,1) & ( 23.77 +   6.08$i$ ) & ( 26.05 +   3.77$i$ ) & ( 24.53 +   3.08$i$ )  \\
(5,2) & ( -8.13 +   4.66$i$ ) & ( -8.84 +   2.55$i$ ) & ( -7.26 +   3.27$i$ )  \\
(5,3) & ( 23.19 +   3.18$i$ ) & ( 20.22 +   4.13$i$ ) & ( 19.95 +   3.95$i$ )  \\
(5,4) & ( -3.77 +   8.91$i$ ) & ( -3.42 +   8.45$i$ ) & ( -2.95 +   8.14$i$ )  \\
\vspace*{2mm}(5,5) & ( 11.21 +  18.55$i$ ) & ( 12.30 +  18.97$i$ ) & ( 12.82 +  20.28$i$ )  \\

(6,0) & (  5.09 +   0.00$i$ ) & (  4.66 +   0.00$i$ ) & (  1.33 +   0.00$i$ ) \\
(6,1) & ( -0.14 +   3.51$i$ ) & (  1.11 +   4.91$i$ ) & (  0.66 +   5.18$i$ ) \\
(6,2) & (  8.72   - 5.38$i$ ) & ( 10.20   - 6.50$i$ ) & ( 10.53   - 6.99$i$ ) \\
(6,3) & ( -4.35 +   1.10$i$ ) & ( -6.06   - 0.12$i$ ) & ( -7.10   - 0.53$i$ ) \\
(6,4) & ( 10.46   - 1.09$i$ ) & ( 11.04   - 0.61$i$ ) & ( 11.10   - 0.39$i$ )  \\
(6,5) & ( -7.08   - 6.34$i$ ) & ( -6.21   - 4.95$i$ ) & ( -5.53   - 3.99$i$ ) \\
(6,6) & (  9.01 +  10.29$i$ ) & (  7.18 +  10.94$i$ ) & (  5.46 +  10.88$i$ )  

\enddata

\end{deluxetable*}

\begin{deluxetable*}{cccccc}
\tablewidth{0pt}
\tablecaption{Low-$\ell$ multipole vector coordinates\label{tab:lowl_vectors}} 
\tablecomments{Low-$\ell$ multipole vectors galactic coordinates
  $(l,b)$ of the first-year \emph{WMAP} data, as computed from the
  spherical harmonics coefficients listed in Table
  \ref{tab:lowl_coeff} by the algorithm of \citet{copi:2004}. 
}
\tablecolumns{4}
\tablehead{ $(\ell, i)$ & LILC & WILC & TOH & TOH-DQ & PE
}
\startdata

$(2, 1)$ & $(130.69^\circ,  13.56^\circ)$ & $(120.95^\circ,  19.79^\circ)$ & $(125.50^\circ,  22.06^\circ)$ & $(118.96^\circ,  25.09^\circ)$ & $(130.83^\circ,  20.83^\circ)$ \\
\vspace*{2mm}$(2, 2)$ & $(  2.52^\circ,  12.71^\circ)$ & $( 15.55^\circ,   3.21^\circ)$ & $(  6.65^\circ,  11.23^\circ)$ & $( 11.15^\circ,  16.62^\circ)$ & $(353.87^\circ,  11.22^\circ)$ \\
$(3, 1)$ & $( 89.22^\circ,  37.70^\circ)$ & $( 95.27^\circ,  37.04^\circ)$ & $( 86.94^\circ,  39.30^\circ)$ & $( 86.94^\circ,  39.30^\circ)$ & $( 85.14^\circ,  36.77^\circ)$ \\
$(3, 2)$ & $( 23.83^\circ,   9.67^\circ)$ & $( 21.73^\circ,   9.39^\circ)$ & $( 22.64^\circ,   9.18^\circ)$ & $( 22.64^\circ,   9.18^\circ)$ & $( 21.65^\circ,  11.36^\circ)$ \\
\vspace*{2mm}$(3, 3)$ & $(312.66^\circ,  10.60^\circ)$ & $(312.98^\circ,  10.71^\circ)$ & $(315.08^\circ,   8.20^\circ)$ & $(315.08^\circ,   8.20^\circ)$ & $(315.31^\circ,   7.52^\circ)$ \\
$(4, 1)$ & $(192.95^\circ,  69.87^\circ)$ & $(199.63^\circ,  70.63^\circ)$ & $(208.64^\circ,  76.73^\circ)$ & $(208.64^\circ,  76.73^\circ)$ & $(150.86^\circ,  74.40^\circ)$ \\
$(4, 2)$ & $(214.60^\circ,  33.56^\circ)$ & $(217.36^\circ,  39.46^\circ)$ & $(206.98^\circ,  31.93^\circ)$ & $(206.98^\circ,  31.93^\circ)$ & $(212.50^\circ,  20.35^\circ)$ \\
$(4, 3)$ & $(333.95^\circ,  28.72^\circ)$ & $(331.69^\circ,  30.23^\circ)$ & $(333.51^\circ,  26.86^\circ)$ & $(333.51^\circ,  26.86^\circ)$ & $(334.43^\circ,  27.64^\circ)$ \\
\vspace*{2mm}$(4, 4)$ & $( 72.41^\circ,   4.30^\circ)$ & $( 71.93^\circ,   6.96^\circ)$ & $( 74.74^\circ,   5.46^\circ)$ & $( 74.74^\circ,   5.46^\circ)$ & $(254.98^\circ,   0.62^\circ)$ \\
$(5, 1)$ & $(227.44^\circ,  56.28^\circ)$ & $(231.38^\circ,  54.54^\circ)$ & $(237.31^\circ,  57.54^\circ)$ & $(237.31^\circ,  57.54^\circ)$ & $(234.76^\circ,  56.33^\circ)$ \\
$(5, 2)$ & $( 97.61^\circ,  37.39^\circ)$ & $(100.79^\circ,  38.52^\circ)$ & $( 98.70^\circ,  38.50^\circ)$ & $( 98.70^\circ,  38.50^\circ)$ & $( 99.99^\circ,  38.77^\circ)$ \\
$(5, 3)$ & $( 43.08^\circ,  36.64^\circ)$ & $( 40.12^\circ,  37.00^\circ)$ & $( 44.67^\circ,  33.54^\circ)$ & $( 44.67^\circ,  33.54^\circ)$ & $( 46.43^\circ,  35.11^\circ)$ \\
$(5, 4)$ & $(288.28^\circ,  31.08^\circ)$ & $(286.47^\circ,  34.23^\circ)$ & $(285.79^\circ,  31.44^\circ)$ & $(285.79^\circ,  31.44^\circ)$ & $(287.88^\circ,  32.28^\circ)$ \\
\vspace*{2mm}$(5, 5)$ & $(177.04^\circ,   1.21^\circ)$ & $(176.05^\circ,   1.20^\circ)$ & $(172.84^\circ,   3.07^\circ)$ & $(172.84^\circ,   3.07^\circ)$ & $(173.91^\circ,   2.23^\circ)$ \\
$(6, 1)$ & $( 30.43^\circ,  52.37^\circ)$ & $( 34.55^\circ,  53.56^\circ)$ & $( 30.66^\circ,  54.88^\circ)$ & $( 30.66^\circ,  54.88^\circ)$ & $( 26.83^\circ,  51.17^\circ)$ \\
$(6, 2)$ & $(242.42^\circ,  52.09^\circ)$ & $(239.66^\circ,  55.52^\circ)$ & $(236.10^\circ,  54.70^\circ)$ & $(236.10^\circ,  54.70^\circ)$ & $(244.92^\circ,  32.48^\circ)$ \\
$(6, 3)$ & $( 86.97^\circ,  32.38^\circ)$ & $( 84.08^\circ,  34.58^\circ)$ & $( 84.62^\circ,  25.31^\circ)$ & $( 84.62^\circ,  25.31^\circ)$ & $( 84.86^\circ,  33.38^\circ)$ \\
$(6, 4)$ & $(282.14^\circ,  24.91^\circ)$ & $(285.49^\circ,  20.96^\circ)$ & $(296.45^\circ,  24.09^\circ)$ & $(296.45^\circ,  24.09^\circ)$ & $(284.29^\circ,  26.57^\circ)$ \\
$(6, 5)$ & $(333.04^\circ,  16.67^\circ)$ & $(337.00^\circ,  17.06^\circ)$ & $(325.01^\circ,  14.58^\circ)$ & $(325.01^\circ,  14.58^\circ)$ & $(330.03^\circ,  15.60^\circ)$ \\
$(6, 6)$ & $(218.37^\circ,   5.48^\circ)$ & $(212.87^\circ,   5.36^\circ)$ & $( 35.65^\circ,   0.60^\circ)$ & $( 35.65^\circ,   0.60^\circ)$ & $(232.36^\circ,  23.45^\circ)$ 

\enddata
\end{deluxetable*}

\end{document}